\documentclass[conference]{IEEEtran}

\usepackage{cite}

\ifCLASSINFOpdf
   \usepackage[pdftex]{graphicx}
\else

\fi

\usepackage{amsmath,amssymb,amsfonts}

\usepackage{algorithmic}

\usepackage{tikz}
\usepackage{textcomp}
\usepackage{lipsum}

\newcommand\copyrighttext{%
  \footnotesize \textcopyright 2018 IEEE. Personal use of this material is permitted.
  Permission from IEEE must be obtained for all other uses, in any current or future 
  media, including reprinting/republishing this material for advertising or promotional 
  purposes, creating new collective works, for resale or redistribution to servers or 
  lists, or reuse of any copyrighted component of this work in other works. 
}

\newcommand\copyrightnotice{%
\begin{tikzpicture}[remember picture,overlay]
\node[anchor=south,yshift=10pt] at (current page.south) {\fbox{\parbox{\dimexpr\textwidth-\fboxsep-\fboxrule\relax}{\copyrighttext}}};
\end{tikzpicture}%
}

\begin{document}
\title{A Method to Derate the Rate-Dependency in the Pass-Band Droop of Comb Decimators}

\author{\IEEEauthorblockN{Ealwan Lee}
\IEEEauthorblockA{GCT Semiconductor, Inc., Seoul, Republic of Korea\\
Email: ewlee@gctsemi.com}}

\maketitle
\copyrightnotice

\begin{abstract}
This paper presents a method to derate the dependency on the decimation factor, $M$, 
of the pass-band droop inherent to $N$-th ordered comb decimators.
It is achieved by cascading a symmetric 3-tap
FIR filter in the integral stage of the corresponding comb decimator and choosing the coefficients only as a function of order $N$.
The proposed derating method derived from the conventional comb decimator can be readily applied to 
any recently developed comb decimator and droop-compensation filter design method.\newline

\end{abstract}


\section{Introduction}
In the design of multi-stage multi-rate filter comprising with
a cascade of FIR filters and down-sampler, comb decimators
have been generally used for the first stage filter
mainly because of its low implementation cost,
over the other interpolation filters~\cite{bib_1,bib_2,bib_3}.
The cost effectiveness of comb decimators has been obtained
with the trade-off of the pass-band spectral flatness, which
had to be compensated by a following so called droop compensation filter.

Although outweighed by its historical popularity and success
in over-sampling analog front-end design such as A/D and 
conversion application~\cite{bib_3,bib_4}, the design problem of
the pass-band droop compensation filter gets more
intricate as the filter order, $N$, increases and the 
decimation factor, $M$, decreases, which is frequently
encountered in recent modern communication system design.
The decimation factor, $M$, has been constantly lowered as the
channel band-width of the communication system gets wider
while the processing clock frequency of the analog front-end is
kept as minimal as possible for low power operation and
technology limit. Meanwhile, the filter order $N$ had to be raised
to meet more tough alias rejection level of the stop band caused
partly by reduced $M$ and partly by the increased order of the
noise-shaping block in the analog front-end.

It is well known that the pass-band flatness of the comb decimator
gets worse as $N$ increases and the stop-band rejection gets
weakened as $M$ decreases. Therefore, if high-order comb decimators are to
be used for a reduced $M$, more sophisticated filters of the
subsequent stages and much care in their design process are required. 
Furthermore, recent communication systems are
forced to support multi symbol rate, thus requiring other stage
filters to be reconfigurable increasing its hardware complexity
compared with that of the fixed tap coefficients.
Correspondingly, many precursory works have been reported to reduce
the required order $N$ with a better pass-band flatness for the same image alias rejection level
\cite{bib_5,bib_6,bib_7} or a simple compensation filters have been proposed to cover both wide range of $N$ and $M$~\cite{bib_8,bib_9,bib_10}.

The main idea of this paper is not dealing with the
problem of compensating overall pass-band droop of the comb decimators.
Instead, we identify that the the pass-band droop can be decomposed into two parts,
 the one dependent only on $N$ and the other one dependent on both $N$ and $M$. 
and we only focus on derating the second part, named as a
pass-band deviation, against $M$ for each $N$. 
It should be reminded that the term of ``derating'' is used exclusively only for the pass-band 
deviation against the change of $M$, as a shortened synonym of desensitization,
while widely accepted terms of ``compensation'' or ``correction'' are used for the overall pass-band droop for clarity in this paper.
Any arbitrary historically well-developed multi-stage multi-rate filter design methods
including the works of~\cite{bib_8,bib_9,bib_10,bib_11,bib_12,bib_13,bib_14} can use the result of this paper 
to derate the dependency on $M$, thus making them more useful in recent development of modern
communication system design.

The rest of our paper is organized like this. At first, the
pass-band deviation, not the overall droop, of the conventional comb decimator 
is addressed and decomposed from the overall pass-band droop in section~\ref{sec_2}. 
Then, we explain how a simple symmetric 3-tap FIR filter
works to derate the dependency of the pass-band deviation to the first order and also
derive the formula for the tap coefficient. Comprehensive
simulation results applied to various comb decimators and compensators
presented in section~\ref{sec_3} and \ref{sec_4} summarize and
confirm the math works obtained in section~\ref{sec_2}, 
followed by the conclusion and suggestion for its use at the end.

\section{Derivation of Derating Filters}
\label{sec_2}
\subsection{Decomposition of the Pass-Band Deviation}
\begin{figure}
\centerline{
\ifCLASSINFOpdf
\includegraphics[width=9cm]{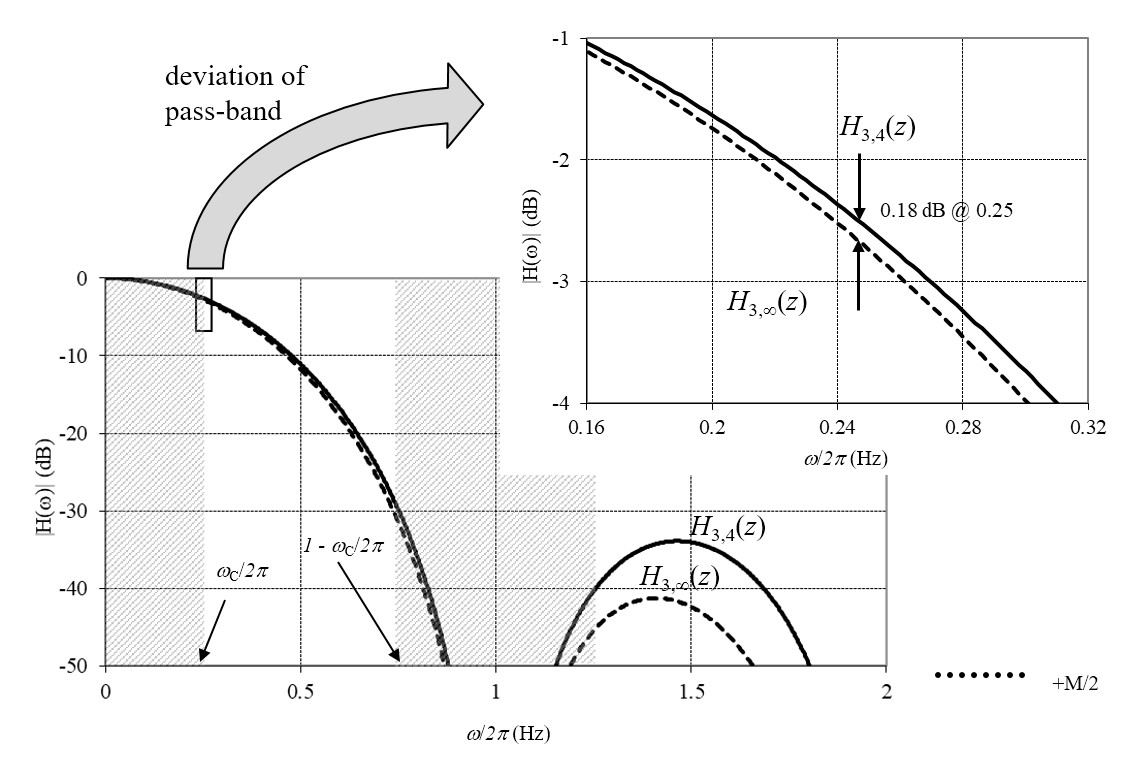}
\fi
}
\caption{Change of  $H_{3,M}(z)$, when $M$ is changed from $\infty$ to 4.}
\label{fig_1}
\end{figure}

To investigate the pass-band deviation of an $N$-th order comb decimator with decimation factor of $M$, 
its transfer function, $H_{N,M}(z)$, has been evaluated at $z = e^{j {\omega / M}}$
for input frequency $\omega$ in the pass band of $[0, \omega_{C}]$ and approximated to
the first order of $(\omega/M)^2$ as formulated in (\ref{eqn_1}). Here, we assume that the output  sampling rate is fixed to $1~Hz$ irrespective of $M$. Therefore, the input frequency, ${\omega} \over {2 \pi} $, is scaled up to $M~Hz$ as $M$ increases.
\begin{IEEEeqnarray}{lCr}
H_{N,M}(e^{j { \omega / M}}) &=&  \left\{ {{ \sin({\omega \over 2}) \over M \cdot \sin({ \omega \over 2M}) }} \right\}^{N}
\nonumber \\
{ } & \approx & {\sin^{N} ({\omega \over 2}) / ({\omega \over 2})^N  \over \left\{ 1 - { 1 \over 3! } \cdot( { \omega \over 2M})^{2}  \right\}^{N}} \nonumber \\
{ } & \approx & {\sin^{N} ({\omega \over 2}) / ({\omega \over 2})^N  \over  1 - { N \over 3! } \cdot ( { \omega \over 2M})^{2} + \Delta }
\label{eqn_1}
\end{IEEEeqnarray}
With this convention, the input signal band-width, or the bandwidth of the decimator, $\omega_{C}$ can also be be fixed as illustrated for $N=3$ in Fig.~\ref{fig_1} against any value of $M$. 
The decimation factor of the following decimation filters, $L$, is chosen typically in the range of $[2 , 4]$.
Thus, the assumption of the worst case of $\omega_{C} = \pi / L \leq \pi /2 $ is acceptable in most cases.
Note that the approximation used in (\ref{eqn_1}) holds valid as long as $(\omega / M) \ll 1$ . 

The approximation obtained from Taylor series expansion has
the terms of only $(\omega /M)^{2 n}$, since $H(e^{j\omega}) = H(e^{- j\omega} )^{*}$.
Consequently, (\ref{eqn_1}) can be decomposed into two : the numerator independent of $M$,
actually the transfer function of the continuous-time $\mbox{sinc}^N$
functions which could be achieved with $M \rightarrow \infty $, and the
denominator accounting for all the dependency on $M$, i.e. the
pass-band deviation. This decomposition shows that the
deviation also becomes worse as $N$ goes high for the same change
of $M$ like the overall droop of the pass-band in the numerator. It is
remarkable that even the low $N$-th order comb decimators, such as
$N=$1 or 2, exhibit a
pass-band deviation proportional to $M^{-2}$ similar to other
high-order case except the difference in multiplication
factors.

\begin{figure}
\centerline{
\ifCLASSINFOpdf
\includegraphics[width=9cm]{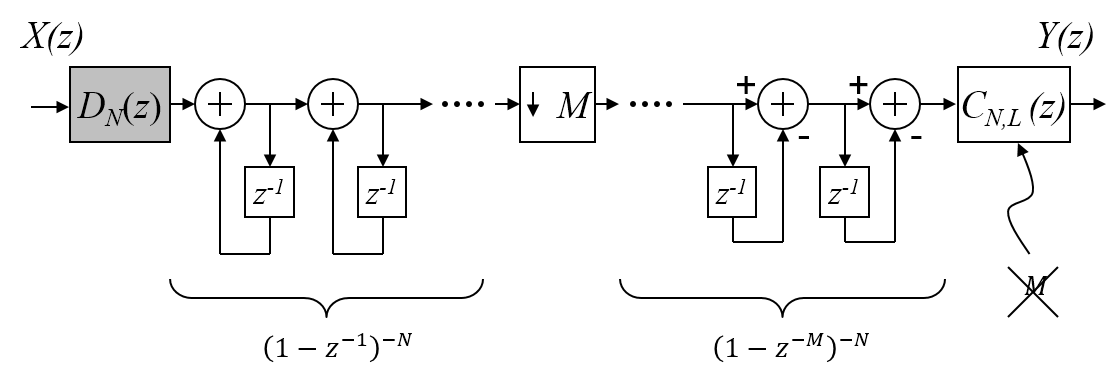}
\fi
}
\caption{The proposed method to derate the dependency of $N$-th order comb decimator on the change of decimation factor of $M$.}
\label{fig_2}
\end{figure}

The denominator in the transfer function of comb decimator represents 
the integral part of the $\mbox{sinc}^N$
filter composed of accumulators in cascade. This fact serves as
a starting point for derating FIR filters
in series of the integral part as shown in Fig.~\ref{fig_2}.

\subsection{First-Order Derating with 3-tap FIR Filters}

The first order dependency on $M^{-2}$ of the denominator in (\ref{eqn_1})
can be cancelled with the following symmetric 3-tap FIR
filter of

\begin{equation}
D_{N}(z) = {1+b_{N} \cdot z^{-1} + z^{-2}  \over 2+b_{N}}
\label{eqn_2}
\end{equation}

, whose first order approximate at $z = e^{j { \omega / M}} $ is

\begin{IEEEeqnarray}{rcl}
D_{N}(e^{j { \omega / M}}) & = &  {e^{-j { \omega / M }}  \over 2+b_{N}} \cdot \left( b_{N}+2 \cdot \cos({ \omega / M})\right) \nonumber \\
{ } & \approx &  {e^{-j { \omega / M }}  \over 2+b_{N}} \cdot \left( b_{N}+2 - { \omega^{2} / M^{2}} + \Delta \right)
\label{eqn_3}
\end{IEEEeqnarray}

It is clear that the convolution of $N$-th order comb decimator and the newly
proposed derating filter results in a new comb-type filter of
\begin{IEEEeqnarray}{rcl}
G_{N,M}(z)  & = & H_{N,M}(z) \cdot D_{N}(z) \nonumber \\
{ } & = & \left( { 1-z^{-M}\over 1-z^{-1}}\right)^{N} \cdot { 1+b_{N} \cdot z^{-1} + z^{-2} \over 2+b_{N}}
\label{eqn_4}
\end{IEEEeqnarray}
, which becomes quite insensitive to the change of $M$ compared to the conventional one since
\begin{IEEEeqnarray}{rcl}
G_{N,M}(e^{j { \omega / M}})  & = & { \sin^{N} ({\omega \over 2})\over ({\omega \over 2})^{N}} \cdot {b_{N} \over 2+b_{N}} \cdot {e^{-j { \omega  / M}}} \cdot \nonumber \\
{ } & { } & \left\{ 1 - \left( { N \omega^2 \over 24M^{2}}\right)^{2}+\Delta \right\}
\label{eqn_5}
\end{IEEEeqnarray}
if $b_{N}$ is chosen to be 
\begin{equation}
b_{N} = 24/N - 2
\label{eqn_6}
\end{equation}
by equating the coefficient of $({\omega / M })^2$ terms in (\ref{eqn_1}) and (\ref{eqn_3})
to derate their effects to the first order approximation.

It is emphasized again that the derating filter, $D_{N}(z)$, is only a function of order $N$ not that of decimation factor, $M$. 
Therefore, only a single 3-tap FIR derating filter exists uniquely for each order of $N$.
On the contrary, the droop compensation filter, $C_{N,L}(z)$ which is typically implemented in the output stage of
comb decimator is a function of $N$ and $L$. Actually, the compensation filter would have been the function of $M$
owing to the pass-band deviation identified in this paper, but its dependency on $M$ is almost removed
by the introduction of $D_{N}(z)$ into integral part.

\subsection{Conditions for the Validity of $D_{N}(z)$}

The approximation and assumptions that we have used in the previous
section limits the range of $N$ for the proposed method to be valid.
Furthermore, it is desired that $| D_{N}(z) | < 1$ to keep
the stop-band rejection of the $G_{N,M}(z)$ not to be degraded from
that of $H_{N,M}(z)$ at least, i.e.
\begin{equation}
| G_{N,M} ( e^{j \omega / M}) | < | H_{N,M} ( e^{j \omega / M}) |
\label{eqn_cond}
\end{equation}
for $ \omega > \omega_{C}$.
 (\ref{eqn_cond}) is satisfied if $b_{N} > 0$, in other words $N < 12$ from (\ref{eqn_6}).
This means that the effectiveness of the proposed
derating method is valid only up to 12 with 3-tap FIR form.
Furthermore, for more rigorous condition of monotonic decreasing property, i.e. no zeros on unit circle except at $\omega = \pm \pi M$
\begin{equation}
|D_{N}(\omega)|= {2 \cdot \cos(\omega / M) + b_{N} \over {2+b_{N}}} \neq 0
\end{equation}
bounds $b_{N} \geq 2$, i.e. $N \leq 6$.
This restriction, however, is
considered to be acceptable covering most practical cases as will be exemplified in the simulation results.

\subsection{Implementation Issues}

\begin{table}

\renewcommand{\arraystretch}{1.3}

\caption{Coefficients and additional word-length for the integer arithmetic implementation of the derating filter.}
\label{tbl_1}
\centerline{
\centering

\begin{tabular}{c|c|c|c}
\hline\hline
$N$ & $b_{N}$ & $A_{N}$ & $W_{b} $\\
\hline
1 & 22 & 1 & 5 \\
2 & 10 & 1 & 4 \\
3 & 6 & 1 & 3 \\
4 & 4 & 1 & 3 \\
5 & 14/5 & 5 & 5 \\
6 & 2 & 1 & 2 \\
\hline\hline
\end{tabular}
\begin{tabular}{c|c|c|c}
\hline\hline
$N$ & $b_{N}$ & $A$ & $W_{b} $\\
\hline
7 & 10/7 & 7 & 5 \\
8 & 1 & 1 & 2 \\
9 & 2/3 & 3 & 3 \\
10 & 2/5 & 5 & 4 \\
11 & 2/11 & 11 & 5 \\
 &  &  &  \\
\hline\hline
\end{tabular}
}
\end{table}

The major merit of the comb decimator
was its low implementation cost resulting from the cascade of accumulators in the renowned recursive form.
However, integer arithmetic is the necessary condition 
for the stability of the cascaded accumulators using modulo arithmetic. 
Thus, it is important to investigate whether the derating filter 
can be implemented in integer arithmetic.

Eq. (\ref{eqn_6}) does not guarantee that $b_{N}$ is always integer for all $N \leq 11$,
which may additionally increase the implementation cost of the 3-tap FIR filter and following
comb decimator. 
Eq. (\ref{eqn_6}), however, assures that $b_{N}$ is a rational number at least. 
So, a proper scaling
with the multiplication of $A_{N} = N /\mbox{GCD}(24,N)$ will make all
its tap coefficients into integer to minimize the extra cost
as summarized in Table~\ref{tbl_1}.
It is noteworthy that the extra output word-length required for the
derating filter is limited by $W_{b} =  \lceil \log_{2}(A_{N} \cdot (2+b_{N})) \rceil$,
which does not grow in proportion to the decimation factor, $M$.

\begin{figure}
\centerline{
\ifCLASSINFOpdf
\includegraphics[width=9cm]{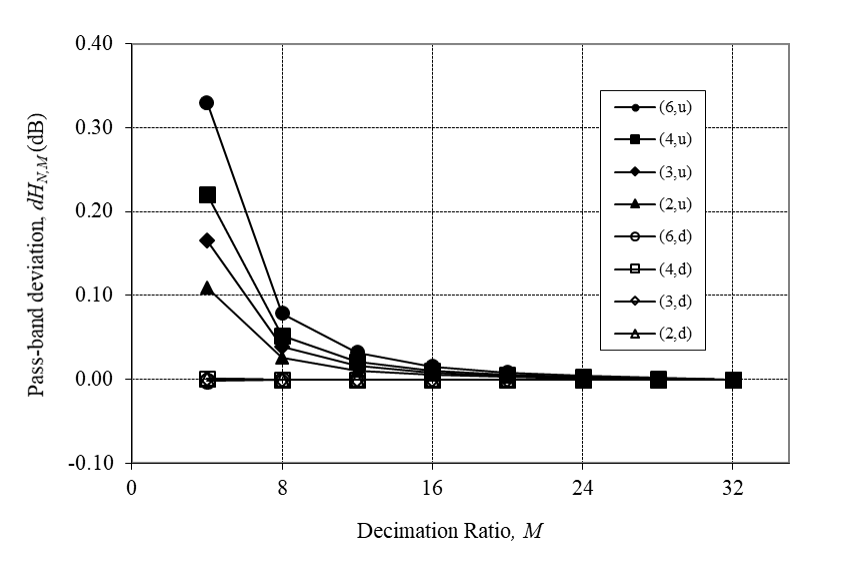}
\fi
}
\caption{Curves of the pass-band deviation as a function of $M$ for the conventional comb decimators with the order of $N=2,3,4,6$. The proposed derating method makes the curve flat compared with the corresponding ones.}
\label{fig_3}
\end{figure}

\section{Simulation Results}
\label{sec_3}
\subsection{Figure of Merits}
The analysis and the condition for the validity of the derived formula
described in the previous section says that linear phase
and monotonic drop in the pass-band is still conserved because
the derating filter is symmetric FIR. Therefore, the evaluation of
$G_{N,M}(z)$ only at the pass band edge, ${\omega / \pi} = 1/L $, 
not all over the pass band, can be used as a measure for the dependency
of $G_{N,M}(z)$ on $M$. It means that a fair comparison
is possible without using any subsequent compensation filters, $C_{N,L}(z)$. 
Furthermore, the measure can be made to represent the pass-band deviation only for each $N$ by
taking the reference value
obtained for a large number of $M$, i.e. 
\begin{subequations}
\begin{equation}
 dG_{N,M} = ||G_{N,M}(e^{j{\pi \over 2 M}}) || - || G_{N,m}(e^{j{\pi \over 2 m}})|_{m \rightarrow \infty} || 
\end{equation}
The same measure can also be obtained for the
the conventional comb decimator, i.e.
\begin{equation}
dH_{N,M} = ||H_{N,M}(e^{j{\pi \over 2 M}}) || - || H_{N,m}(e^{j{\pi \over 2 m}})|_{m \rightarrow \infty} ||
\end{equation}
\label{eqn_7}
\end{subequations}

\subsection{Case of Conventional Comb Decimators}
From (\ref{eqn_1}), it is conjectured that the pass-band deviation is always
positive, i.e. $dH_{N,M} > 0$ as shown in the graph of Fig.~\ref{fig_3}.
For brevity, the curves for un-derated and
derated comb decimators of $N=\{2,3,4,6\}$ only are put together in Fig.~\ref{fig_3}.

The un-derated conventional ones exhibit noticeable dependency of pass-band deviation on $M$ as $N$ increases,
as predicted in (\ref{eqn_1}).
The range of $M$ taken for simulations is from 4 to 32 with the
step size of 4 while $L$ is fixed to 2 for worst cases, 
which is expected to cover most practical cases.
The derated comb decimators labeled as ($N$, d) with
$N=\{2,3,4,6\}$ in the graph show their superiority to the
conventional ones labeled as ($N$, u).

\section{Extension to Recent Works}
\label{sec_4}
 Although derating filter, $D_{N}(z)$, has been derived from the coventional comb decimator, 
 the application is not confined to it.
It can be extended to most of the modified comb decimators and pass-band droop compensators 
published in recent works as exemplified in the following sub-sections.

\subsection{Comb Decimators with a Filter Sharpening}
\label{sec_4a}
Filter sharpening method has been used to improve both the pass-band droop and the stop-band rejection
of the conventional comb decimator\cite{bib_6,bib_7}.
In most cases, the lowest order sharpening is widely adopted as in
\begin{equation}
F_{N,M}(z) = 3 \cdot H_{N,M}^{2}(z) - 2 \cdot H_{N,M}^{3}(z).
\end{equation}
\begin{figure}
\centerline{
\ifCLASSINFOpdf
\includegraphics[width=9cm]{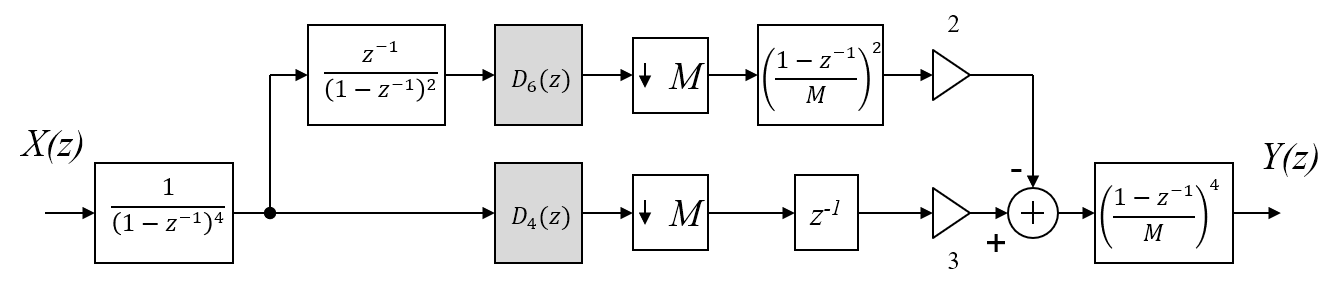}
\fi
}
\caption{Example of the sharpened comb decimators with the derating filter.}
\label{fig_4}
\end{figure}
\begin{figure}
\centerline{
\ifCLASSINFOpdf
\includegraphics[width=9cm]{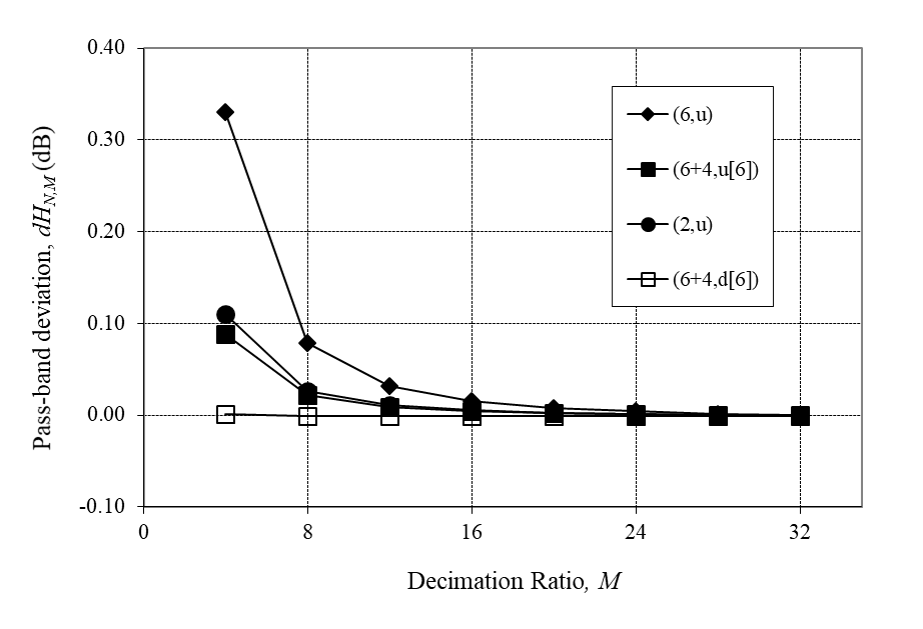}
\fi
}
\caption{Curves of the pass-band deviation for the sharpened comb decimators.}
\label{fig_5}
\end{figure}

The order of the base comb decimator should be even number to align the group delay
of the two terms for sharpening thus limiting the lowest order of $H_{N,M}(z)$ to 2,
which means
\begin{subequations}
\begin{equation}
H(z) = \left( {1\over M} \cdot {  1-z^{-M} \over 1-z^{-1}} \right)^{2} 
\end{equation}
\begin{equation}
F(z) = 
	3z^{-M} \cdot \left( { 1-z^{-M} \over 1-z^{-1} } \right)^{4}
	- {2z^{-1} \over M^{2}} \cdot \left( { 1-z^{-M}\over 1-z^{-1}} \right)^{6}
\end{equation}
\end{subequations}
Here, introduction of either $z^{-M}$ or $z^{-1}$ is for the rate invariant alignment of two components in time domain.
The pass-band droop of this filter sharpening method for the equivalent rejection level is much smaller
than that of the corresponding conventional comb decimator. Hence it is very advantageous
for the design of the following pass-band droop compensation filter in differential stage.

Each term of $H(z)^2$ and $H(z)^3$ can be derated by $D_{4}(z)$ and $D_{6}(z)$ respectively as shown in Fig.~\ref{fig_5}. The position of the derating filter can be either the input or output of the integral part. The dependency of the resulting derated decimator $F(z)$ becomes
quite flat against the change of the decimation factor, $M$, as exemplified in the graph of Fig.~\ref{fig_5}.

\subsection{Comb Decimators with Distributed Zeros in Stop-Band}
\label{sec_4b}
\begin{figure}
\centerline{
\ifCLASSINFOpdf
\includegraphics[width=9cm]{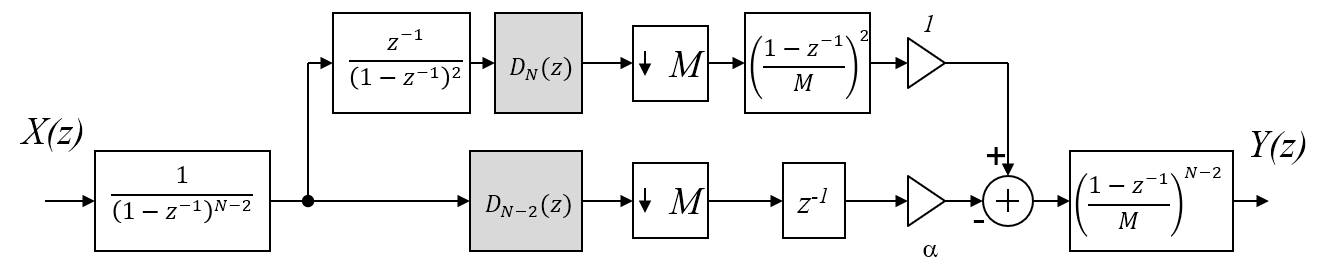}
\fi
}
\caption{Derated second($N=2$) and third($N=3$) order comb decimators with bifurcate zeros.}
\label{fig_6}
\end{figure}

\begin{figure}
\centerline{
\ifCLASSINFOpdf
\includegraphics[width=9cm]{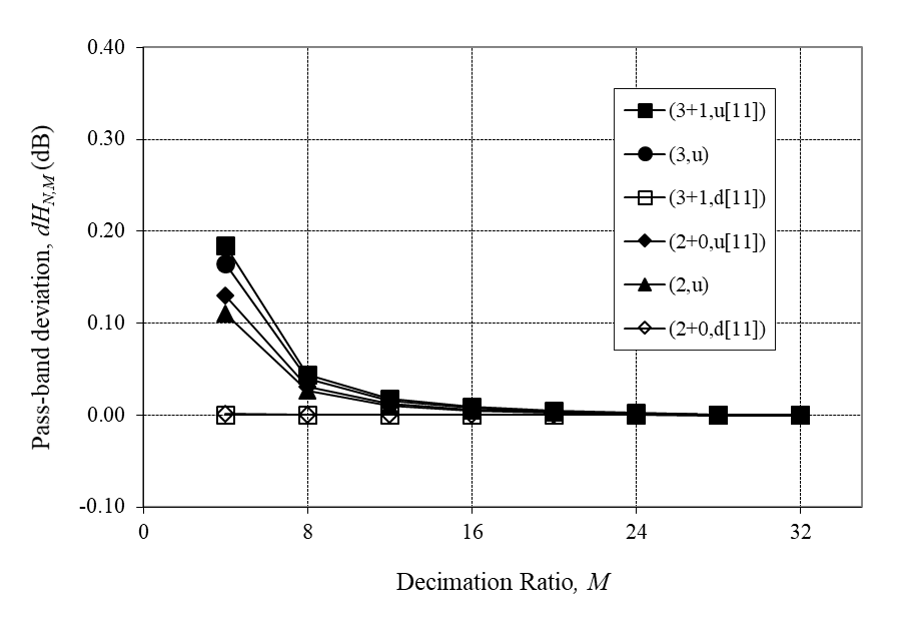}
\fi
}
\caption{Curves of the pass-band deviation for the second and third order comb decimators with bifurcate zeros.}
\label{fig_7}
\end{figure}

Lots of structures have been proposed to improve the alias rejection level in stop-band especially folding into the desired band. Among them, \cite{bib_11} explains clearly how the multiple zeros at folding frequency of $f = n~Hz$ bifurcate although the structure itself is a degenerate case of \cite{bib_11} as the authors acknowledged.

It may be hard to tell the difference of the modified comb decimators between Fig.~\ref{fig_4} and Fig.~\ref{fig_6},
but thorough investigation of those twos reveals the followings :
Filter sharpening technique finds the uniquely defined coefficients and ratio between the higher and lower order comb decimator for a given sharpening order
and does not move the zeros of the folding band 
collapsing into multiple zeros as in conventional comb decimator.
The other works focusing more on improving the alias rejection level in the folding band
move, distribute or bifurcate the multiple zeros in the folding band as outlined in~\cite{bib_11}.
Relieved from the constraint, the order of the outer comb decimator can be drastically reduced from 4 to 1 for $N=3$ as an example.
Consequently, $D_{3}(z)$ and $D_{1}(z)$ are used in replace of $D_{6}(z)$ and $D_{4}(z)$ 

Our proposed derating method still works as shown inf Fig.~\ref{fig_7}.
Note that the curve of underated \cite{bib_11}, (3+1, u\cite{bib_11}), is slighltly above that of (3, u) since the modified comb decimator is a small perturbation of conventional filter unlike filter sharpening.

This method can be easily applied to the second order comb decimator replacing $D_{3}(z)$ and $D_{1}(z)$ with $D_{2}(z)$ and $D_{0}(z)$, respectively. $D_{0}(z)=z^{-1}$ meaning $b_{0}~\rightarrow~\infty$ although it is not listed in table~\ref{tbl_1} for its peculiar form.

\subsection{Effects on the Maximally Flat Droop Compensator}
\label{sec_4c}
\begin{figure}
\centerline{
\ifCLASSINFOpdf
\includegraphics[width=9cm]{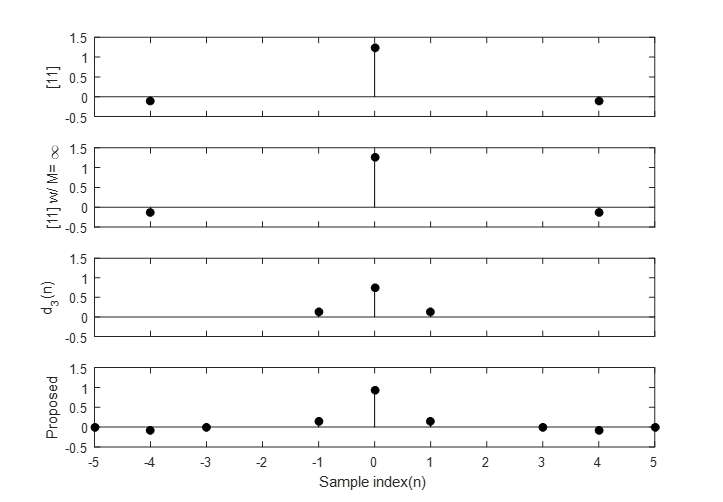}
\fi
}
\caption{Comparison between single stage maximal flat 3-tap FIR compensator and two stage version 
with a proposed derating filter for $(N,M) = (3,4)$.}
\label{fig_8}
\end{figure}

Unlike other computer-aided design methodology, there have been works on the pass-band droop compensator 
with the coefficients in closed mathematical form under the maximal flatness condition of the pass band\cite{bib_13, bib_14}. Especially, \cite{bib_13} derived an explicit scalar closed-form formula for the coefficents of 3 and 5-tap FIR filter respectively applicable to either narrow or wide band depending on $L$. 

It is noted that our proposed derating filter makes the coefficients of this compensation filter, $C_{N,L}(z^{M})=c_{0}+c_{1}z^{-M} + c_{2}z^{-2M}$, 3-tap FIR for narrow band case with $L>4$, independent of $M$ and simplified to
\begin{IEEEeqnarray}{rcl}
c_{0} (= c_{2}) & = & \left.- { N \over 32 } \cdot { 1 - M^{-2} \over 1 - 2^{-2}} \right|_{M\rightarrow \infty} \rightarrow - \frac{N}{24} \nonumber \\
c_{1} & = & 1 - (c_{0} + c_{2}).
\end{IEEEeqnarray}
A similar technique can be readily applied to the wide band case just by letting $M \rightarrow \infty $ for the 5-tap FIR filter coefficients.

This result leads us to get a more hardware-efficient two-stage multiplierless implementation of the maximally flat compensator without any more concern about the dependency on the decimation factor, $M$
, unlike the previous single-stage implementation as compared in Fig.~\ref{fig_8}.

All the flimsy terms dependent on $M^{-2}$ are removed from the compensator coefficients with extra pre-emphais in differential stage. A derating filter in the integral stage, essentially an LPF, nullifies thus simplifies the coefficients of overall filters both derating filter and the pass-band droop compensator. It is also evident that the word-length of compensator is no more dependent on the decimation factor, $M$.

\newpage

\section{Conclusion}
We addressed the dependency of the pass-band droop in $N$-th order comb decimators on its decimation
factor, $M$, and proposed a proactive method adopting an
extra low-complexity FIR filter to derate it. A group of 3-tap low-pass FIR filters
in the integral part of the comb decimators
has shown to reduce the deviation of the compensated pass-band
significantly while keeping the other properties almost intact.
It is noted that it holds not only to the conventional comb decimators but also to any 
recently studied comb decimators and pass-band droop compensators.

Besides, a pretty accurate cost increase model still preserving the modulo integer arithmetic has been suggested
for the quantitative analysis of the additional implementation cost caused by the adoption of the derating filter.
Consequently, the proposed method can be readily applied
to comb decimators studied so far and any other uncharted comb-type decimators 
if their pass-band deviation against
the change of decimation factor, $M$, is to be minimized.

\end{document}